# Probing the magnetic state by linear and non linear ac magnetic susceptibility measurements in under doped manganite $Nd_{0.8}Sr_{0.2}MnO_3$

S. Kundu[a] and T. K. Nath[*]

*Department of Physics and Meteorology, Indian Institute of Technology, Kharagpur, West Bengal 721302, India*
[a]*email:* souravphy@gmail.com

***Abstract***

We have thoroughly investigated the entire magnetic states of under doped ferromagnetic insulating manganite $Nd_{0.8}Sr_{0.2}MnO_3$ through temperature dependent linear and non linear complex ac magnetic susceptibility measurements. This ferromagnetic insulating manganite is found to have frequency independent ferromagnetic to paramagnetic transition temperature at around 140 K. At around 90 K ($\approx T_f$) the sample shows a second frequency dependent re - entrant magnetic transition as explored through complex ac susceptibility measurements. Non linear ac susceptibility measurements (higher harmonics of ac susceptibility) have also been performed (with and without the superposition of a dc magnetic field) to further investigate the origin of this frequency dependence (dynamic behavior at this re – entrant magnetic transition). Divergence of 3rd order susceptibility in the limit of zero exciting field indicates a spin glass like freezing phenomena. However, large value of spin relaxation time ($\tau_0 = 10^{-8}$ s) and small value of coercivity (~22 Oe) obtained at low temperature (below $T_f$) from critical slowing down model and dc magnetic measurements, respectively, are in contrast with what generally observed in a canonical spin glass ($\tau_0 = 10^{-12} - 10^{-14}$ s and very large value of coercivity below freezing temperature). We have attributed our observation to the formation of finite size ferromagnetic clusters which are formed as consequence of intrinsic separation and undergo cluster glass like freezing below certain temperature in this under doped manganite. The results are supported by the electronic - and magneto - transport data.

---

*Corresponding author: tnath@phy.iitkgp.ernet.in

Tel: +91-3222-283862
Area code: 721302,
INDIA

## 1. Introduction

The perovskite rare earth manganites with general formula $R_{1-x}A_xMnO_3$ (R= rare earth elements, A= divalent elements) have been studied quite intensely from the discovery of their colossal magnetoresistance (CMR) property as this proved a potential way of application of such materials in the form of magnetic field sensors and memory devices[1]. As a matter of fact, it is not only their potential for application but also the fascinating physics that these manganites involve made them an interesting family of material for scientific study. It is already known that manganites are strongly correlated material having a close interplay between their spin, charge, orbital and lattice degrees of freedom [2,3]. Due to this strong correlation, manganites show some unusual electronic and magnetic properties. Around an optimal doping level ($x \approx 0.3$) the double exchange mechanism plays the key role behind the observation of ferromagnetism and CMR property. It is this CMR property and ferromagnetism which attracted most of the research interest on optimally doped manganites [4]. The half doped manganites have also been studied quite intensively due to their interesting property of simultaneous charge and orbital ordering [5]. The overdoped and underdoped manganites are comparatively less explored by the physicists. In fact, due to the dominance of antiferromagnetic superexchange interaction competing with double exchange interaction and strong Jahn - Teller interaction results a complex magnetic state in these systems. The detailed understandings of the underlying physics of the complex magnetic state could be interesting in such manganites.We have investigated the magnetic property of under doped manganite $Nd_{0.8}Sr_{0.2}MnO_3$ through linear and non linear ac susceptibility measurements. According to the bulk reported phase diagram this system shows a ferromagnetic insulating to paramagnetic insulating phase transition around 150 K [1]. Our susceptibility and resistivity data show the same features. Moreover, we have observed a second frequency dependent magnetic transition at lower temperature originating most likely from cluster glass type freezing. In some Mn site doped manganites such frequency dependent re - entrant transition were observed and mostly of them were claimed to have glassy character [6-8]. Some rare earth cobaltites are also known to have glassy character [9, 10]. The glassyness in these cases was attributed to the intrinsic inhomogenity and phase separation in the samples. It has also been observed that A-site disorder has a major role to play in the observed glassy behavior in some manganites [11]. However, in an otherwise undoped manganite like ours, such reentrant frequency dependent transition is not common in literature. N. Rama *et al.* [12] have reported similar magnetic transition in $Pr_{0.8}Sr_{0.2}MnO_3$. In this investigation, we have reported for the first time to the best of our knowledge, a detailed linear and nonlinear complex ac susceptibility study along with dc field superposition effect and explored the cluster glass like freezing of ferromagnetic clusters in $Nd_{0.8}Sr_{0.2}MnO_3$ sample.

## 2. Experimental details

The manganites $Nd_{0.8}Sr_{0.2}MnO_3$ was synthesized employing chemical pyrophoric reaction route. Requisite amount of TEA (Triethanolamine) was mixed with the stoichiometric solution of high purity $Nd_2O_3$, $Sr(NO3)_2$, and $Mn(CH_3COO)_2$. The solution was then heated at 180 °C and stirred continuously. The finally obtained black fluppy powder was ground and sintered at 1150 °C in air for 8 hrs to produce the polycrystalline manganites of $Nd_{0.8}Sr_{0.2}MnO_3$.

The XRD (x – ray diffraction) pattern shown in Fig. 1(a) confirms the perovskite structure with no observable impurity phase present. From FESEM (field emission scanning electron microscopy) image (Fig. 1(b)) it is evident that the sample is composed of nearly micrometer size particles.

The complex ac magnetic susceptibility measurement (both linear and non linear) on this sample was done by employing an indigenously developed ac susceptibility set up by us. In this set up the sample was placed inside one of the two oppositely wound pickup coils while the excitation ac magnetic field (h) was applied by a solenoid coaxial with the pickup coils. The signal was measured employing a lock - in - amplifier (model-SR830). We have also measured the higher order non linear ac magnetic susceptibility just by detecting the signal at frequencies integral multiple of the exciting frequency. The temperature was monitored by a temperature controller (Lakeshore, model-325). A calibrated cernox censor (Lakeshore made) was used to sense the temperature with 0.05 K resolutions. The dc magnetization was measured employing a homemade vibrating sample magnetometer having sensitivity better than $10^{-3}$ emu. Similar kind of electronic equipments that were used in the ac susceptibility set up were also employed for detecting the signal and monitoring the temperature in the magnetometer.

The magnetization m of a specimen can be expressed, when excited by a small ac magnetic field h, as

$m = m_0 + \chi_1 h + \chi_2 h^2 + \chi_3 h^3 + \ldots$ ($m_0$ is spontaneous magnetization) [13]. (1)

Here, $\chi_1$ is the linear susceptibility and $\chi_2$, $\chi_3$ etc. are the higher order or the nonlinear susceptibilities. For $h = h_0 \sin(\omega t)$, the induced voltage in the secondary pick up coils can be written as [10],

$$V = -K\omega h_0[(\chi_1 + \frac{3}{4}\chi_3 h_0^2 + \frac{5}{8}\chi_5 h_0^4 + \ldots)\sin\omega t + (\chi_2 h_0 + \chi_4 h_0^3 + \frac{15}{16}\chi_6 h_0^5 + \ldots)\sin 2\omega t$$
$$-(\frac{3}{4}\chi_3 h_0^2 + \frac{15}{16}\chi_5 h_0^4 + \ldots)\sin 3\omega t + \ldots] \quad (2)$$

Here, K is the constant of the set up which is to be determined by calibration of the set up with a known sample. We have used the high moment paramagnetic salt $Gd_2O_3$ to calibrate our system.

In the low field limit (small $h_0$) the series in the first brackets rapidly converge. So the different terms in equation (2), can be written as,

$$|V_\omega| = K\omega|\chi_1|h_0 \quad (3)$$

$$|V_{2\omega}| = K\omega|\chi_2|h_0^2 \quad (4)$$

$$|V_{3\omega}| = K\frac{3}{4}\omega|\chi_3|h_0^3 \quad (5)$$

$V_\omega$, $V_{2\omega}$, $V_{3\omega}$, are the measured voltages by the lock in amplifier at frequencies ω, 2ω and 3ω.

By properly manipulating the voltages and knowing the calibration constant K the linear and the non linear susceptibilities can be determined. Each of these susceptibilities has a real and imaginary component. For example, the linear susceptibility $\chi_1$ can be expressed as,

$$\chi_1 = \chi_1^R - j\chi_1^I \tag{6}$$

$\chi_1^R$ is the real component and $\chi_1^I$ is the imaginary component of the linear susceptibility. With a lock in amplifier both the components can be measured by measuring the signal in phase with the exciting field (imaginary part) and out of phase with the exciting field (real part). It is to be noted that a small enough ac magnetic field (h) is required as mentioned above, to make all the approximation applicable. Moreover, high exciting field can destroy any critical features to be observed [14], especially in measuring the higher order susceptibilities. Application of high field can itself smear out the magnetic phase transition and contributions from domain wall movement/rotation may appear. Very low ac magnetic field enables us to measure the true spin susceptibility of a magnetic system. The limit of this field is again determined by the sensitivity of the set up itself. For small applied field the signal strength also becomes weak and so a compromisation has to be made. In our case the applied ac field (h) is always kept at a very small value (below 7 Oe rms).

### 3. Results and discussion

To probe the magnetic behavior and investigate the relaxation process of the system we have performed ac susceptibility measurements over a wide range of frequency (three decades). The measured real part of linear ac susceptibility of our sample ($\chi_1^R$) at different frequencies at a probing field of h = 2 Oe is shown in the Fig. 2. The inset of Fig. 2 shows the imaginary part which represents the loss or dissipation in the sample. The temperature variation of real part shows two transitions, with a high temperature transition occurring around 140 K and a low temperature cusp ($T_f$) around 90 K. The sharp drop of susceptibility ($\chi_1^R$) at around 140 K can be attributed to the ferromagnetic (FM) to paramagnetic (PM) phase transition. The bulk phase diagram also shows a FM to PM transition almost at the same temperature. At around 90 K susceptibility decreases rapidly with the decrease of temperature showing a second re - entrant transition around this temperature. The imaginary part of ac susceptibility ($\chi_1^I$) shows two distinct peaks at these two transition region. The height of the low temperature peak is greater than that at the higher temperature. This clearly indicates that the loss in the system is quite high around this low temperature transition region compared to that in the FM-PM transition. The presence of an appreciable peak in $\chi_1^I$ around the second transition at the low temperature region indicates that this transition is a genuine magnetic transition and not experimental artifacts. Most interestingly, the magnitude of the imaginary part is non zero in the FM region. It is quite evident that while the high temperature FM-PM transition is not frequency dependent, the transition (cusp) at the lower temperature is strongly frequency dependent. This transition temperature shifts towards the higher temperature with the increase of frequency. The magnitude of susceptibility is slightly suppressed with the increase of frequency at this transition region and the susceptibility curves become separated from each other completely at the lower temperatures. The peak of the $\chi_1^I$ of the sample increases in magnitude with the increase of frequency and show a shift to higher temperature side with the increase of frequency (not shown). All the features described above regarding the second transition at lower temperature seem similar to that observed in case of a freezing process of a spin glass around the freezing temperature. However, we want to mention here that the frequency dependence of susceptibility is not unique to a spin freezing process and can also be observed in a superparamagnet and even in complex ferromagnetic systems.

In order to probe the complex magnetic state in the system, the ac susceptibility with a superimposed dc magnetic field has also been investigated in details. If a dc magnetic field $H_{dc}$ is applied along with the ac field, all the susceptibilities (linear and non linear) will be modified. From Fig. 3 we can observe that with the increase of the magnitude of the dc field the FM-PM transition is suppressed considerably. With the increase of dc field the transition shifts to lower temperature with an overall suppression of $\chi_1^R$ in the ferromagnetic regime. It is generally observed for a common ferromagnet that the susceptibility maxima (Hopkinson peak) just below the FM-PM transition is slightly suppressed on application of dc magnetic field [15,16]. In our case we have observed an enhanced suppression of $\chi_1^R$ over a wide temperature range on application of a small dc magnetic field (45 Oe). It is also observed that the low temperature transition is only slightly suppressed compared to the FM- PM transition region on application of a maximum of 45 Oe dc field whereas, at the lowest temperature regime the susceptibility is almost independent of dc field.

To further investigate the magnetic state of our sample we have studied the non linear or the higher harmonic response of the ac susceptibility which are much more sensitive around the phase transitions compared to its linear counterpart and more effective in some cases to analyze the complex magnetic system. The even harmonic $\chi_2$ can be observed if there is a presence of spontaneous magnetization or a symmetry breaking internal field [15]. The odd harmonic $\chi_3$ has been efficiently used to characterize a spin glass or a superparamagnet and also to differentiate between them. The 3$^{rd}$ harmonic ($\chi_3$) shows a very sharp peak at the freezing and a broad peak at the blocking temperature of a spin glass and superparamagnet, respectively [17]. For a spin glass $\chi_3$ is divergent in nature in the limit h→0 and $f$→0 around the freezing temperature. For other magnetic systems diverging nature is not observed around magnetic transitions. In this way, measurement of $\chi_3$ gives a high level confidence in determining glassyness in any magnetic specimen.

The measured real part of 2nd harmonic of susceptibility ($\chi_2^R$) of $Nd_{0.8}Sr_{0.2}MnO_3$ without and with the application of dc field ($H_{dc}$), is shown in the inset of Fig.3. Temperature variation of $\chi_2^R$ shows a sharp peak at the FM-PM transition region.This is due to the onset of spontaneous magnetization in the ferromagnetic regime. On application of a small superimposed dc field the peak height of $\chi_2^R$ increases. However, we note that the peak height (at 140 K) shows a non – monotonic dependence on $H_{dc}$ in the dc field range of $0 < H_{dc} < 15$ Oe. Further increase of $H_{dc}$ suppresses the peak. Peak position is found to shift towards the lower temperature with the increase of dc field. The 3rd harmonic susceptibility ($\chi_3^R$) shows a broad peak at the low temperature transition region (T ~ 90 K) as evident from Fig. 4. However, the measured 3rd harmonic show a divergent nature as the magnitude of the ac field (h) is decreased towards zero. The peak values of $\chi_3^R$ ($|\chi_3^R|_{max}$) are plotted as a function of h to clarify this behavior (Inset of Fig. 4). It is clear that, the value of ($|\chi_3^R|_{max}$) is diverging in the low field regime in the limit of h → 0 whereas, it shows a saturating nature as the field is increased. This indicates that the magnetic phase transition at the low temperature region occurs due to cooperative freezing phenomena, weather it constitutes a classical spin glass or the behavior is due to dipolar interaction among the magnetic clusters. The log – log plot of $\chi_3^R$ vs. reduced temperature (T-T*)/T* (T* is the peak temperature of $\chi_3^R$) is shown in the inset of Fig. 4. A saturating nature of the quantity log ($\chi_3^R$) is clear as T approaches T*. This behavior is a characteristic of a spin glass system [18]. It was experimentally found that the application of dc magnetic field (~ 45 Oe) does not have any effect on $\chi_3$ around the low temperature re-entrant transition region.

The dc magnetization of the sample was recorded at 500 Oe field and shown in the inset of Fig. 5. Magnetization of the sample initially rises with temperature and attains a maximum at around 73 K. With further increase of temperature magnetization of the sample decreases slowly and follows a sharp fall at around 140 K. The transition at 140 K is indicative of the FM-PM transition as obtained from the ac susceptibility data. The magnetic hysteresis (M vs. H) of the sample was also registered at different temperatures in the low field regime. Fig. 5 shows the M vs. H curves at different temperatures. Maximum dc field applied was 500 Oe. The magnetization of the sample at all temperatures show a non saturating behavior as the applied field is very small. Even with the application of a small field, one can observe a clear and feasible evolution in the M-H behavior of the sample with the change of temperature. The maximum value of magnetization has gradually decreased with the increase of temperature and the M-H curve becomes almost linear near the FM-PM transition temperature. Our main objective of this measurement was to have an idea about the coercivity of the sample, an important issue in systems where glassy behavior is evident. One noticeable fact is that the coercivity ($H_C$) of the sample is very small (~22 Oe) even at the lowest measured temperature. Large value of coercivity (even a few hundreds of Oe) is hallmark of a spin glass system. In this regard, our system does not match with a canonical spin glass. Inset of Fig. 5 shows the variation of $H_C$ with temperature. $H_C$ is highest at the lowest measured temperature and falls sharply at around 90 K and then decreases slowly with the increase of temperature up to the FM-PM transition region where it becomes almost zero.

We have also performed the electronic transport study of the sample with and without the application of external magnetic field to support our magnetic data. The resistivity vs. temperature plot is shown in Fig. 6. The sample is insulating in the whole measured temperature range. Moreover, resistivity increases very sharply just below 100 K. We note that this is the temperature regime (around 90 K) where the ac susceptibility data show a transition and frequency dependence. On application of a 5 T magnetic field a pronounced suppression in the resistivity can be observed just below 90 K. The magnetoresistance (MR) vs. T data (Inset of Fig. 6) clearly indicates this suppression by showing a minimum at this temperature. A small kink is observable at around 140 K, which is most likely due to the enhanced intrinsic MR around Curie temperature.

We have carefully analyzed our experimental observation as follows. We have first employed the thermal activation model and analyzed the low temperature frequency dependent transition by applying the Arrhenius law given by $\tau = \tau_0 \exp (E_a/k_B T_f)$ [19]. Here, $E_a$ is the activation energy. The freezing temperature $T_f$ is taken as the temperature corresponding to the peak in $\chi_1^I$ in the low temperature region as the position of the exact observed transition point from $\chi_1^R$ is not prominent. $\tau$ is taken as the experimental time scale $1/f$. We first note that the variation of $T_f$ is not linear in $f$ as evident from the inset of Fig. 7. The variation of $T_f$ with frequency is very sharp in the low frequency regime whereas, a slow variation $T_f$ is observable in the high frequency regime. To verify the validity of the Arrhenius law in this system ln ($\tau$) is plotted as a function of $T_f$. The best fit results of this plot give $\tau_0 = 10^{-50}$ s and $E_a = 1224000$ K. Both these values are unphysical. So, we can rule out the possibility of any thermally activated process and hence the possibility of a superparamagnetic blocking at this temperature regime. Moreover, the M vs. H/T plots at different temperatures (not shown) do not overlap and indicates that the frequency dependence of susceptibility is not due to blocking of magnetic moments. The large grain size (micrometer order) of this sample also indicates that superparamagnetic behavior is not favorable in this manganite. Another way of analyzing the frequency dependent peak in ac-$\chi$ is calculating the shift in peak with decade change in frequency. The quantity $\Delta T_f / T_f \Delta(\log\omega)$ gives a value of 0.06. This value is well within the regime observed for insulating spin glasses [19]. In case of a spin glass the frequency dependence is very often analyzed by employing the conventional critical slowing down model. According to this theory the dynamic scaling law of the form $\tau = \tau_0 (T_f/T_g-1)^{-z\nu}$ is very commonly used [19, 20]. Here, $\tau_0$ is the spin-relaxation time, $z\nu$ is the critical exponent and $T_g$ is the spin glass transition temperature. The fitting procedure in this case remains the same. The fitting of this equation to our experimental data (Fig. 7) is quite good with the obtained fitting parameters given by, $T_g$ = 73 K, $z\nu$ = 9 and $\tau_0 = 10^{-8}$ sec. For a conventional spin glass system typical value of $\tau_0$ is of the order of $10^{-12}$ - $10^{-14}$ s. We have obtained a higher value of $\tau_0$ compared to that of a conventional spin glass system whereas, the value of $z\nu$ is well within the regime of a spin glass [19]. Interestingly, we note that the value of $T_g$ ($\approx$ 73 K) is the temperature corresponding to the peak position of the dc magnetization vs. temperature plot. In fact, $T_g$ is equivalent to the freezing temperature of a spin glass in the limit of zero frequency. The low temperature transition is very similar to a transition to a cluster glass phase as the obtained value of the characteristic time is quite higher compared to that of a spin glass. A high value of spin relaxation time was obtained by B. Ray *et al.* [21] in $Nd_{0.7}Sr_{0.3}MnO_3$ and the phase was designated as a cluster glass. The peak around $T_f$ in the temperature dependent $\chi_3^R$ arises due to this cluster glass type freezing process. The peak is broad in nature. For a canonical spin glass a very sharp divergence at the freezing temperature can be observed. The dc field dependence of ac - $\chi$ is quite remarkable in our system. The FM-PM transition is found to be suppressed on application of dc magnetic field whereas, at low temperature transition and below that the system is very robust against dc field. For a conventional spin glass system the sharp susceptibility peak around the freezing temperature is strongly suppressed and gets rounded off on application of dc magnetic field [19]. The broad peak in $\chi_3$ around the freezing temperature, robustness of the low temperature transition against dc magnetic field together with the obtained high value of relaxation time and low value of coercivity differentiates our system from a conventional spin glass. The peak temperature in the $\chi_2$ can be taken as the exact value of the FM-PM transition temperature or $T_C$, the Curie temperature of the system. Reduction of $T_C$ and suppression of peak height in $\chi_2$ on application of dc field directly indicates the reduction of FM exchange interaction and hence the spontaneous magnetization in the system under dc field.

The behavior of the system can be understood with the concept of the formation of finite size ferromagnetic clusters and interaction among them. The formation of such short range ferromagnetism is a consequence of phase separation phenomena which is very commonly observed in manganites. The coexistence of large carrier-rich FM clusters with smaller carrier-poor FM clusters has been reported earlier [10,22]. This is indirectly indicating the so called electronic phase separation in manganites. The larger clusters will, understandably, favor ferromagnetism in the sample. Whereas, according to their hypothesis, smaller ones will enhance glassyness. We also propose that the ferromagnetic region of the sample does not have long ranged order but formed of small clusters of ferromagnet. Formation of FM clusters in manganites has been reported earlier. The formation of FM cluster in $La_{0.85}Ca_{0.15}Mn_{0.95}Fe_{0.05}O_3$ was attributed to the random field created by the PM clusters present in the ferromagnetic regime by H.U. Khan *et al* [6]. In $La_{0.5}Gd_{0.2}Sr_{0.3}MnO_3$ formation of FM clusters were identified by P. Dey *et al* [23]. The AFM interaction between Gd and Mn moments was thought to have promoted the formation of FM clusters. In our system, it is quite understandable that AFM interaction in the $Mn^{3+}$-O-$Mn^{3+}$ will be appreciable beside double exchange due to the abundance of $Mn^{3+}$ ions in this system. This AFM interaction resists the formation of long ranged ferromagnetic order throughout the system but only allows the formation of finite size FM clusters. The ferromagnetic clusters may coexist with the non ferromagnetic regions and the whole system can be thought of a magnetically phase segregated state of a manganite. This kind of phase separation in terms of existence of ferromagnetic regions inside antiferromagnetic matrix was theoretically predicted by M Yu Kagan *et al. in* low doped manganites [24]. Rivadulla *et al.* [25] experimentally showed the possibility of formation of such a state in manganites where frustration and glassyness may arise due to intercluster interaction. They argued that finite size of the ferromagnetic clusters which restrains the divergence of the magnetic correlation length $\xi$, had a role to play in the system behavior. They also showed that dc field can modulate the cluster size/concentration which in turn modifies the frustration and glassyness in the system. From the experimental evidences, it seems that a dc field induced size modulation of the ferromagnetic clusters can take place in our system. On application of dc field, probably cluster size decreases and limits the value of $\xi$ and hence the spontaneous magnetization of the system. This is reflected in the suppression of $T_C$ and absolute value of $\chi_2$ and $\chi_1$ under dc field. When the applied dc field is very low a slight enhancement of $\chi_2$ as well as $\chi_1$ can be observed. We believe that this enhancement is due to the contribution from the higher order susceptibility [13] and not a genuine property of the system. This can be understood by the fact that $T_C$ decreases monotonically with the increase of dc field although the absolute value of susceptibility initially increases and again decrease at higher applied dc field. At the low temperature regime the ferromagnetic clusters undergo random dipolar interaction between them.

This causes a frustration and collective relaxation in the system as manifested by the frequency dependence of ac-$\chi$ .The system undergoes a cluster glass type freezing due to this dipolar interaction. It is known that an increase in cluster number or cluster concentration will enhance the random dipolar interaction [26]. As a result the frustration leading to glassy behavior will be enhanced. Probably, there is a reduction of cluster size and hence an increase in cluster concentration with the decrease of temperature occurs in our system. Around a certain temperature (90 K) we observe the glassy behavior. The enhanced frustration at around 90 K is reflected in the sharp rise in resistivity at the same temperature regime. Double exchange, which is enhanced by the alignment of spins or ferromagnetic moments, will be hindered at this temperature regime due to increased frustration and disorder in this system. At the low temperature regime dc field cannot further reduce the cluster size which is already small. This could be the reason for the observed robustness of the system in the low temperature regime against the application of dc field. The MR of such manganite sample is known to originate from spin polarized inter - granular tunneling phenomenon and double exchange mechanism (intrinsic MR). Intrinsic MR is appreciable only around $T_C$. The most significant spin polarized tunneling MR ($MR_{SPT}$) is proportional to $M^2$ and 1/T [27]. The increase in negative MR with the decrease of temperature is qualitatively consistent with the dc magnetization data (magnetization increases with the decrease of temperature). We observe that there is a sharp decrease in the MR with the decrease of temperature in the lowest temperature region. The freezing of the spins/ clusters might be the reason of this reduced MR.

Lastly, we want to mention the possibility of an intra cluster contribution leading to glassyness which may originate from random ferromagnetic and anti ferromagnetic bonds in each cluster of this sample. We believe that system having random interaction within a single cluster would have more similarities with a real spin glass and would give rise to more spin glass like properties to be observed in this system. Our present study shows that this system displays many properties which are not commonly observed in a canonical spin glass. So, inter cluster interaction seems the most plausible explanation for our observation of glassy behavior in the system.

**Conclusions:**

In summary, we can say that we have investigated in details the magnetic properties of under doped manganite $Nd_{0.8}Sr_{0.2}MnO_3$ employing ac magnetic susceptibility measurement techniques. Besides having a ferromagnetic to paramagnetic phase transition we have observed another magnetic transition in the low temperature region. This transition seems to be originating from the freezing of ferromagnetic clusters. Around this transition point ac susceptibility shows frequency dependence and 3$^{rd}$ harmonic of susceptibility shows a broad peak and diverges in the zero field limit. This divergence of $\chi_3$ is a typical signature of a spin glass. However, we have observed very low value of coercivity of the sample from M-H curves and large value of spin relaxation time ($10^{-8}$ s). These observations are not in accord with genuine glassy behavior. Our observation has been attributed to inter cluster dipolar interactions. It seems that the phase separation and inter cluster interaction is responsible for the observation of some glass like behaviors in the system. Resistivity and magneto transport measurements supports our observed magnetic properties.

## Acknowledgement

One of the authors (T. K. Nath) would like to acknowledge the financial assistance of Department of Science of Technology (DST), New Delhi, India through project no. IR/S2/PU-04/2006.

**Figure captions**

**Fig. 1.** (a) XRD micrograph of the sample. (b) FESEM image of the sample.

**Fig. 2.** Variation of the real part of linear ac susceptibility ($\chi_1^R$) with temperature at different frequencies (h = 2 Oe). The inset shows the plot of imaginary part of ac susceptibility ($\chi_1^I$) measured at $f$ = 999.3 Hz and h = 2 Oe.

**Fig. 3.** (Color online) The variation of the real part of linear ac susceptibility under different dc magnetic fields. The inset shows the plot of the real part of 2$^{nd}$ harmonic of ac susceptibility ($\chi_2^R$) under different dc magnetic fields.

**Fig. 4.** (Color online) Plot of 3rd harmonic of ac susceptibility ($\chi_3^R$) measured at different ac magnetic fields. The upper left inset shows the peak value of $\chi_3^R$ as function of ac magnetic fields (h). The upper right inset shows the log - log plot of $\chi_3^R$ as a function of reduced temperature (T-T*)/T*.

**Fig. 5.** (colour online) The M-H behavior of the sample at different temperatures. The upper left inset shows the variation of coercivity with temperature. The bottom right inset shows the variation of magnetization of the sample with temperature recorded at 500 Oe field.

**Fig. 6.** Measured resistivity of the sample as a function of temperature without the application of magnetic field and under 5 T field. Inset shows the plot of magnetoresistance as a function of temperature.

**Fig. 7.** Plot of log (1/$f$) as a function of freezing temperature $T_f$. Inset shows the variation of $T_f$ with frequency.

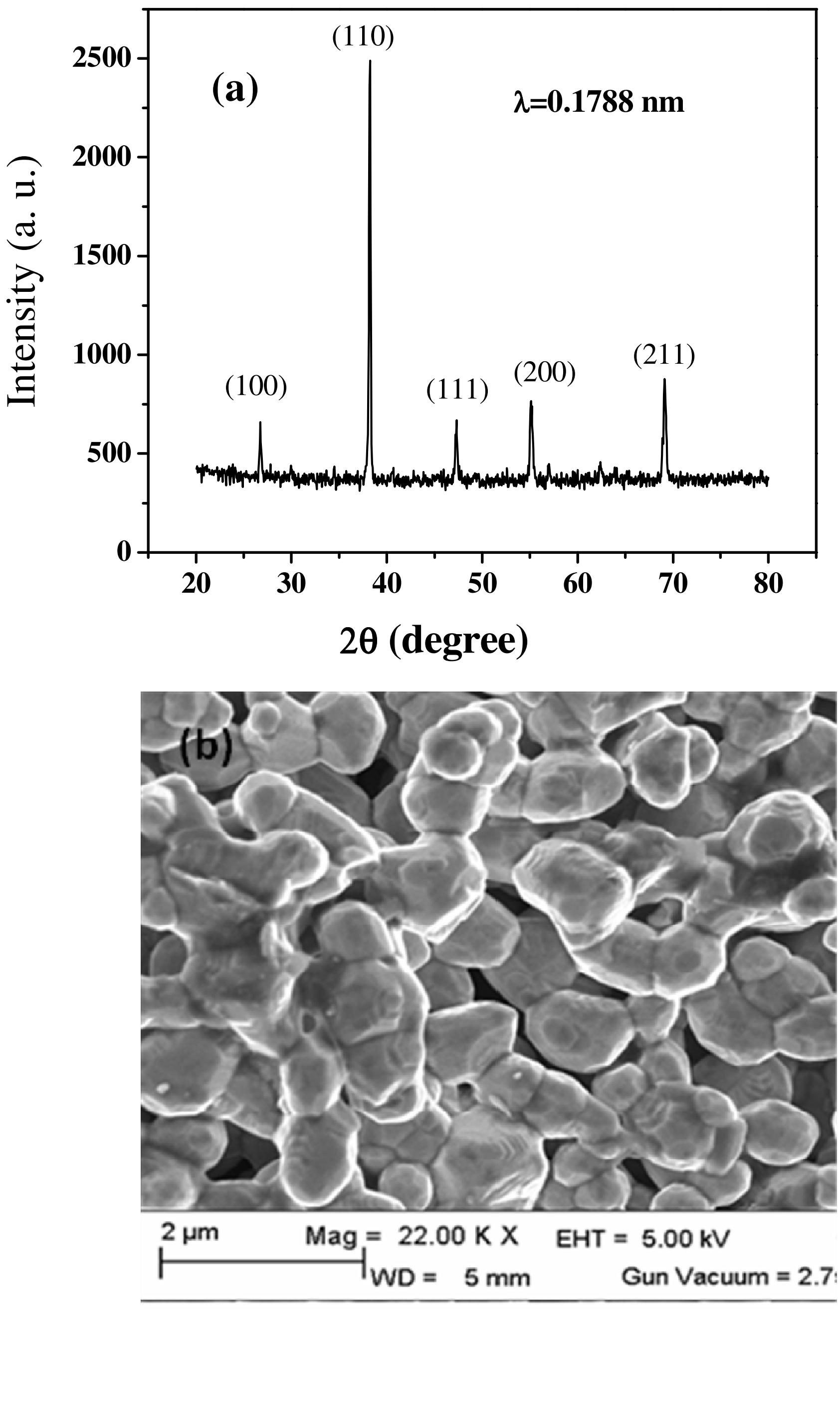

**Figure 1**

**S. Kundu *et al.***

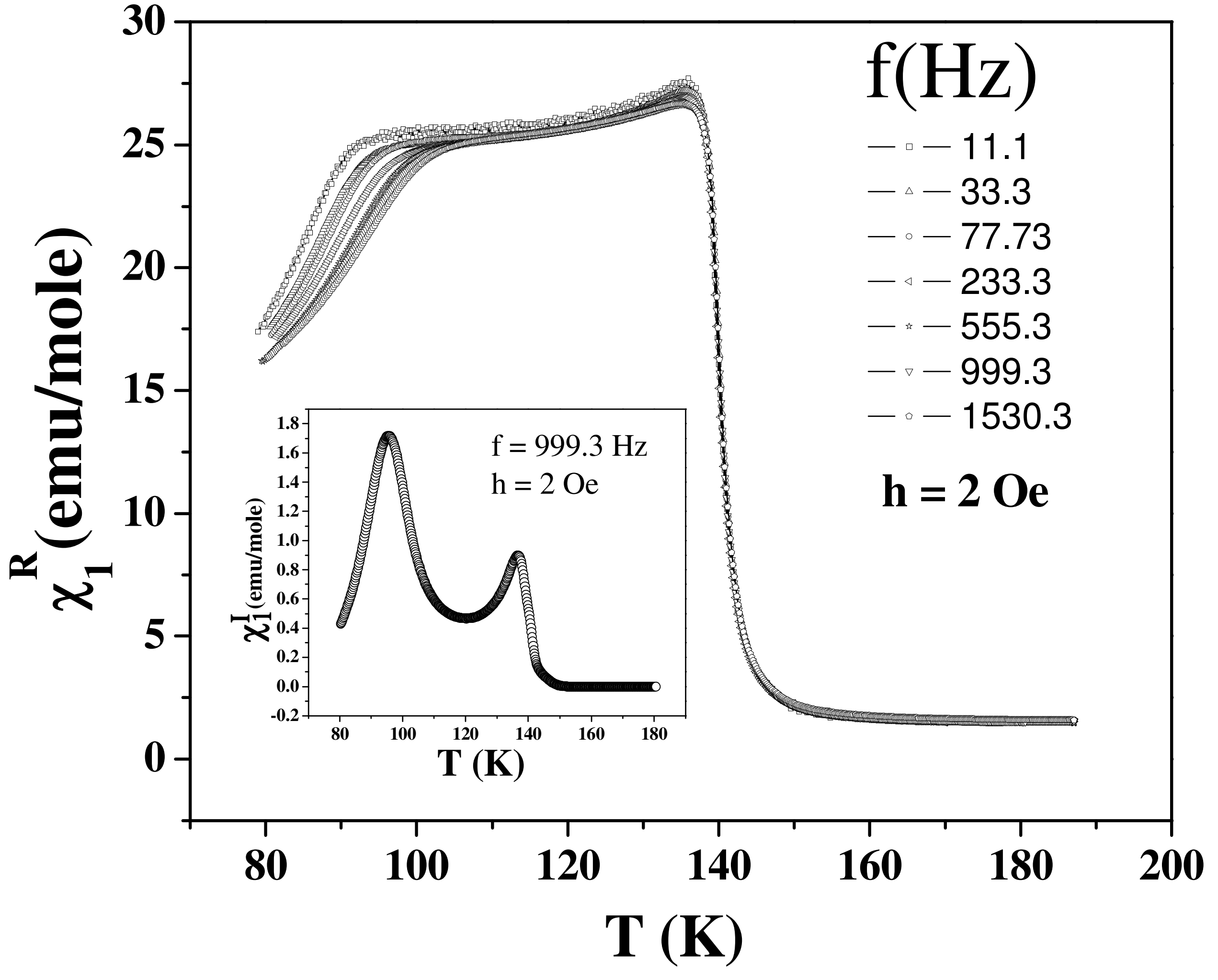

**Figure 2**

**S. Kundu *et al.***

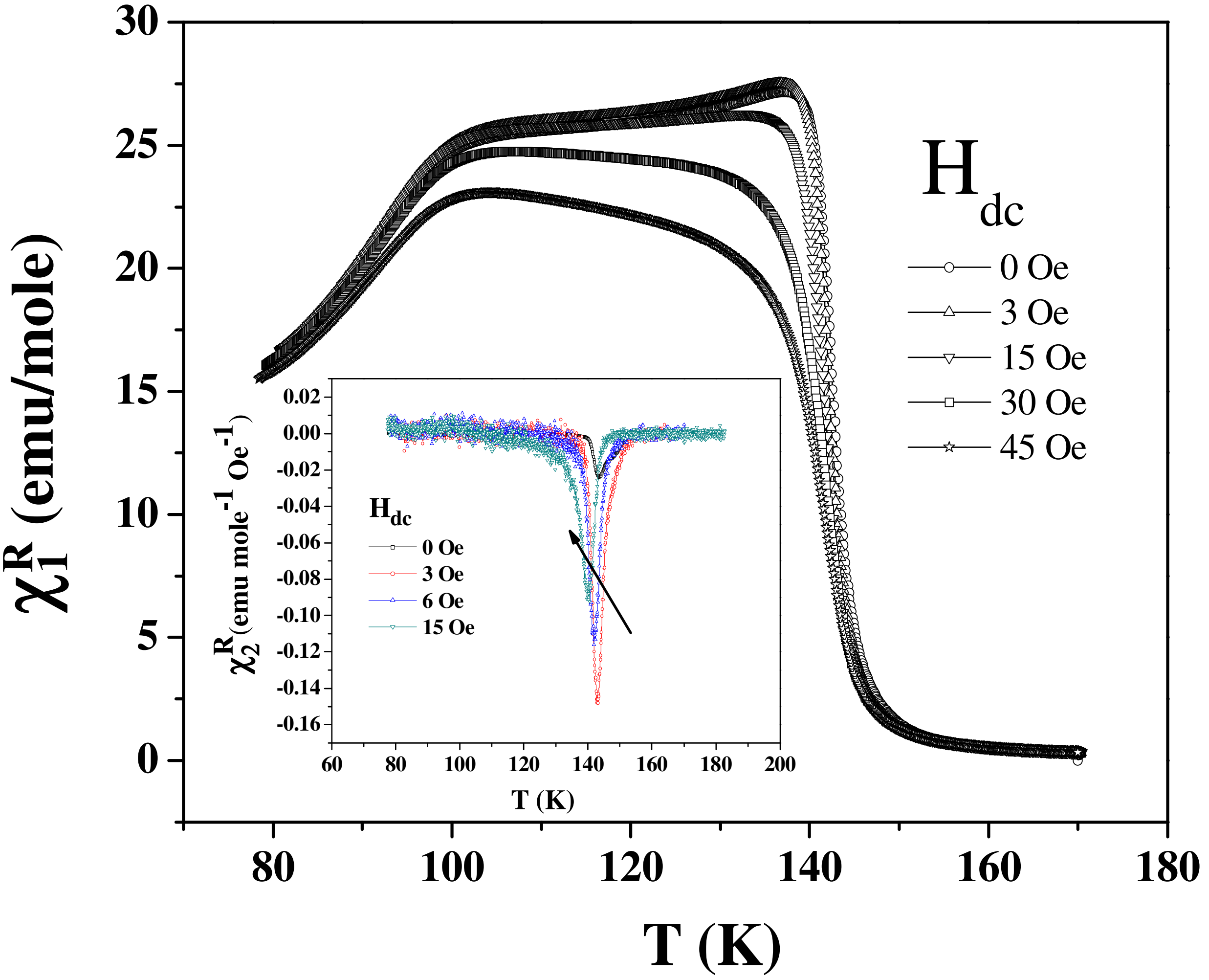

**Figure 3**

**S. Kundu *et al.***

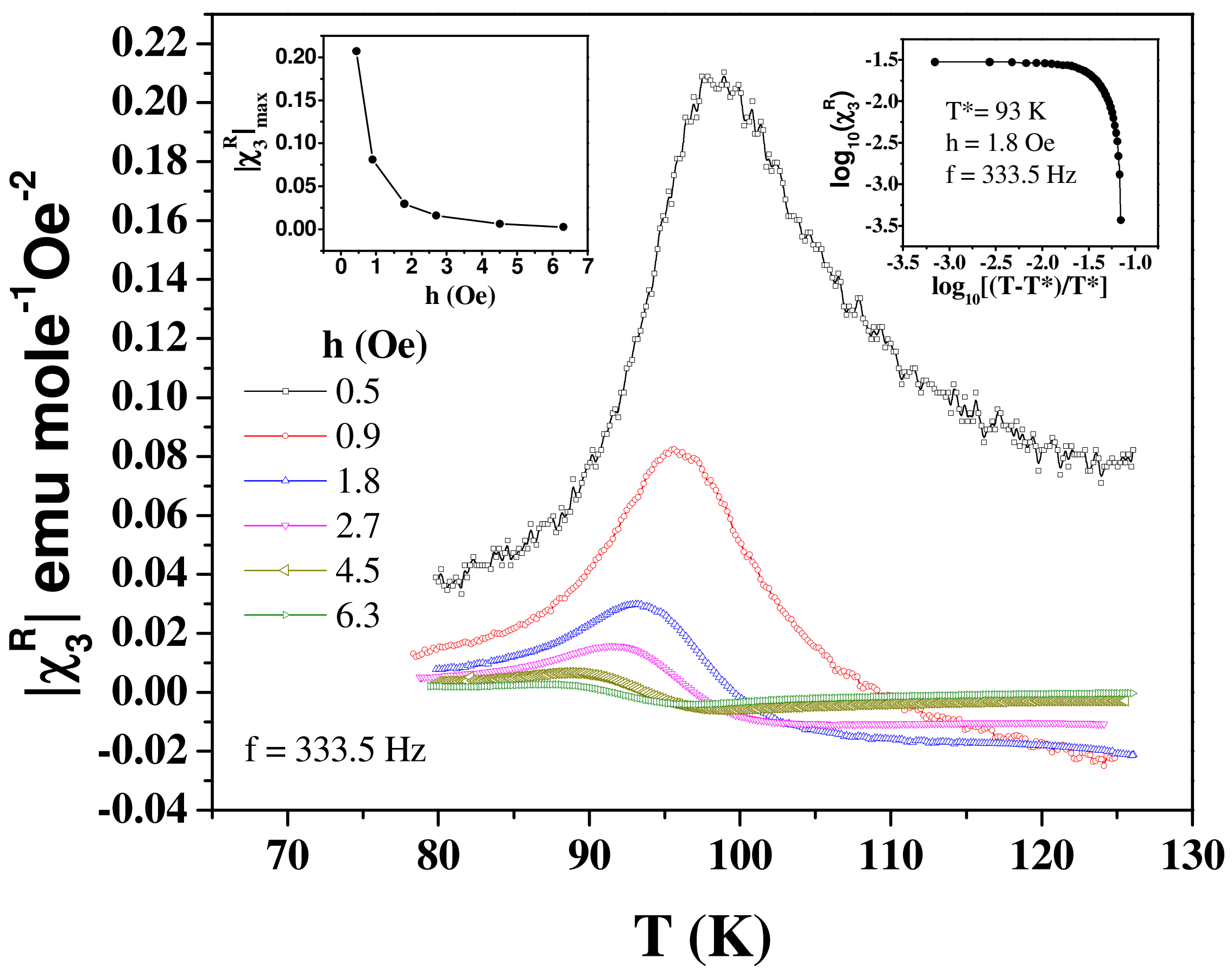

**Figure 4**

**S. Kundu *et al.***

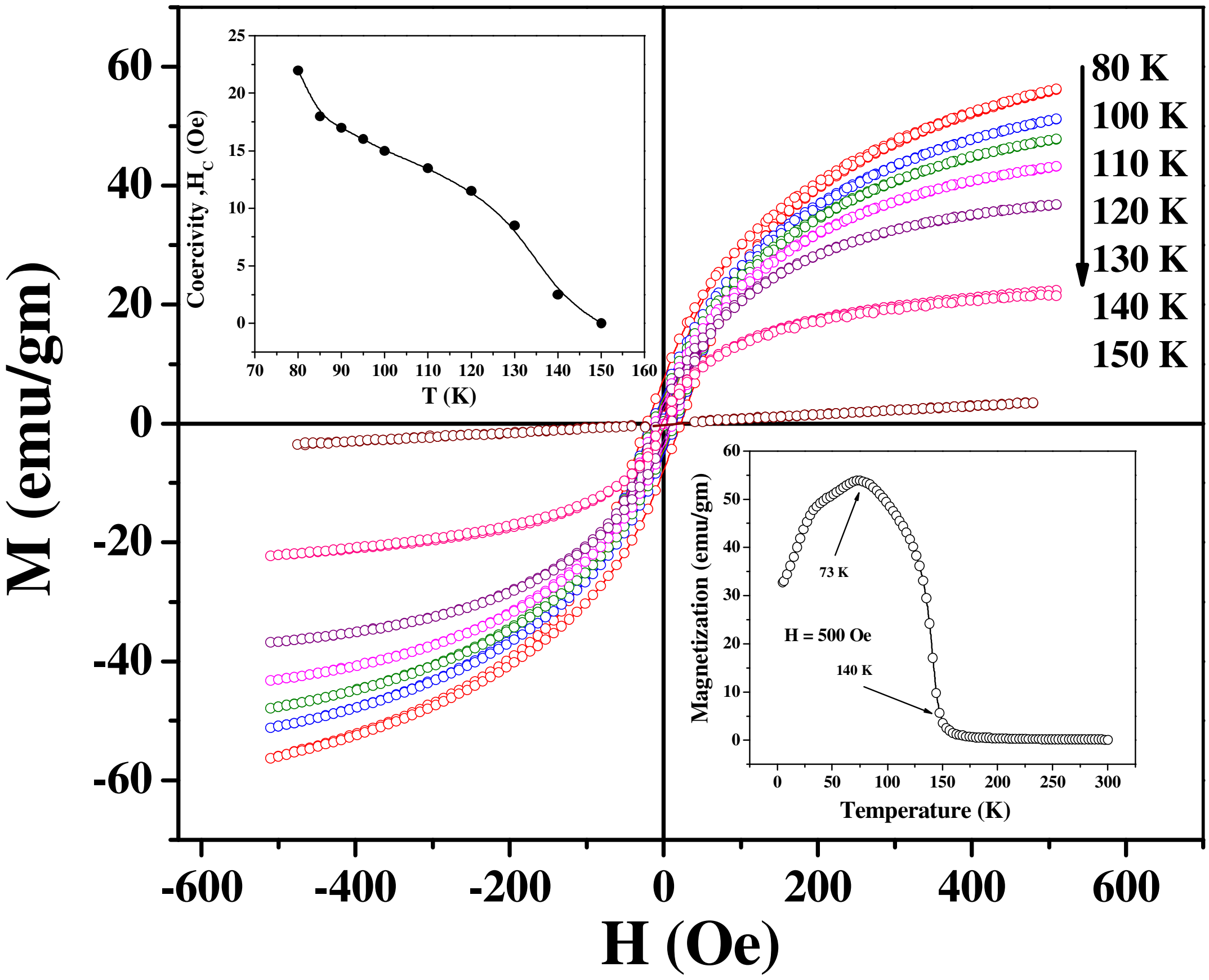

**Figure 5**

**S. Kundu *et al.***

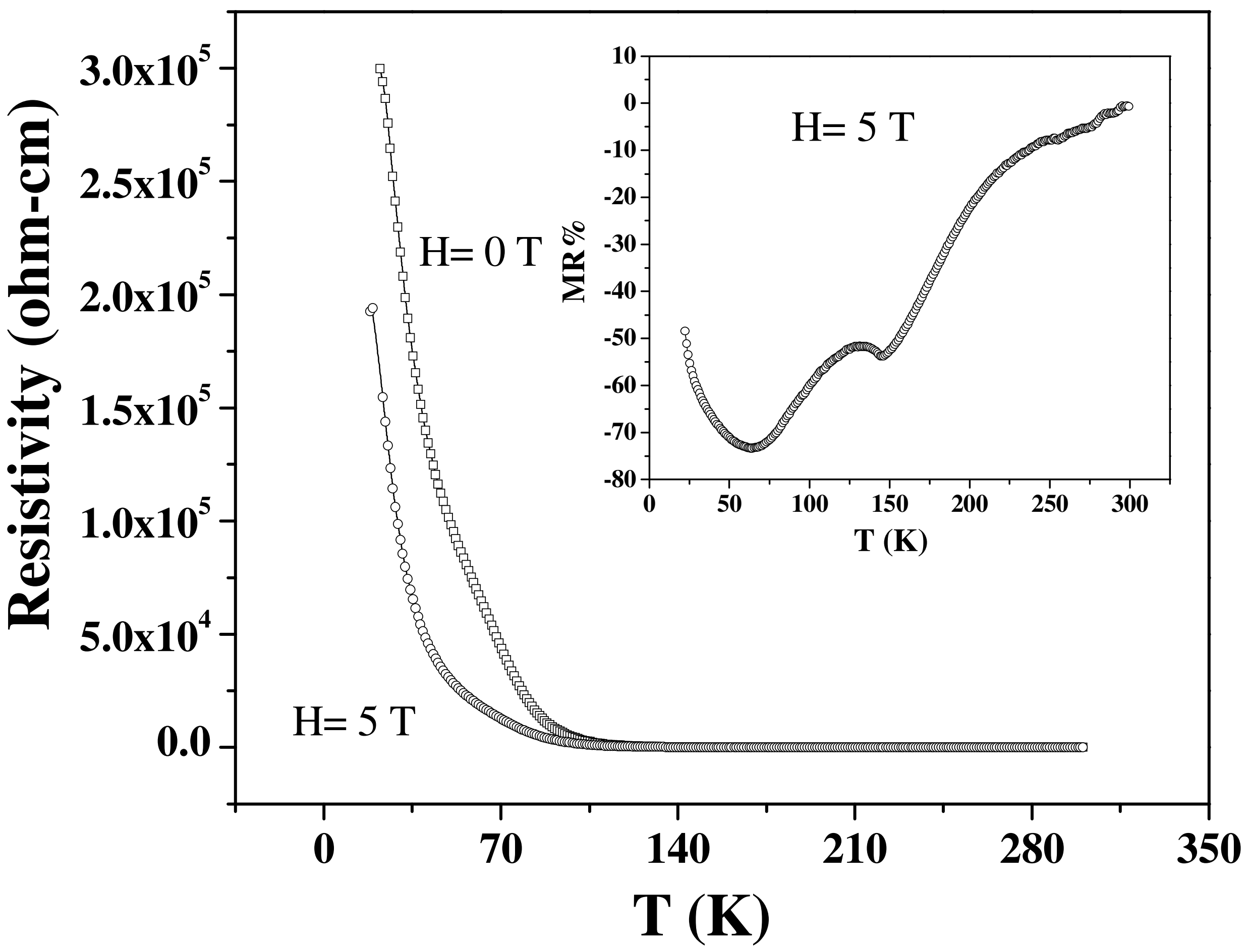

**Figure 6**

**S. Kundu *et al.***

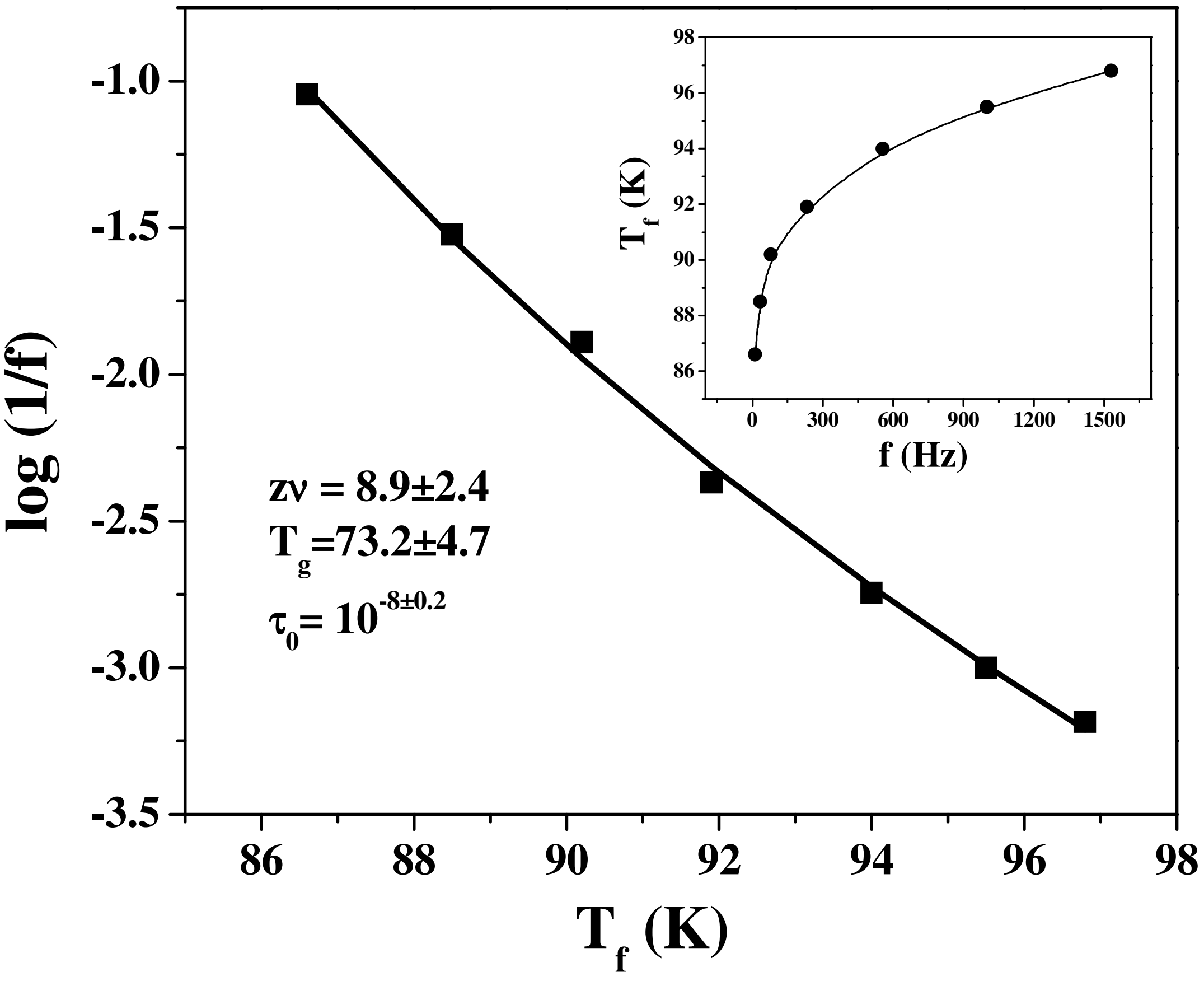

**Figure 7**

**S. Kundu *et al.***